\documentclass[12pt]{iopart}
\usepackage{iopams}
\usepackage{epsfig}

\begin{document}

\title[Phase diagrams of CeRhIn$_5$ and CeCoIn$_5$] {High pressure phase diagrams of CeRhIn$_5$ and CeCoIn$_5$
 studied by ac calorimetry}

\author{G Knebel\dag, M-A M\'{e}asson\dag, B Salce\dag, D Aoki\dag, D
Braithwaite\dag, J P Brison\ddag\ and J Flouquet\dag\S }

\address{\dag\ D\'{e}partement de Recherche Fondamentale sur la Mati\`{e}re Condens\'{e}e,
SPSMS, CEA Grenoble, 38054 Grenoble Cedex 9, France}

\address{\ddag\ Centre de Recherche sur les Tr\`{e}s Basses Temp\'{e}ratures, CNRS, 38042 Grenoble Cedex 9, France}

\address{\S\ Institut de Physique de la Mati\`{e}re Condens\'{e}e de Grenoble, 38042 Grenoble Cedex 9, France}

\begin{abstract}
The pressure-temperature phase diagrams of the heavy fermion antiferromagnet CeRhIn$_5$ and the
heavy fermion superconductor CeCoIn$_5$ have been studied under hydrostatic pressure by ac
calorimetry and ac susceptibility measurements using diamond anvil cells with argon as pressure
medium. In CeRhIn$_5$, the use of a highly hydrostatic pressure transmitting medium allows for a
clean simultaneous determination by a bulk probe of the antiferromagnetic and superconducting
transitions. We compare our new phase diagram  with the previous ones, discuss the nature (first or
second order) of the various lines, and the coexistence of antiferromagnetic order and superconductivity. The link between the collaps of the superconducting heat anomaly and the broadening of the antiferromagnetic transition points to an inhomogeneous appearence of superconductivity below $P_c \approx 1.95$~GPa. Homogeneous bulk superconductivity is only observed above this critical pressure. We present a detailed analysis of the influence of pressure inomogeneities on the specific heat anomalies which emphasizes that the observed broadening of the transitions near $P_c$ is connected with the first order transition. For CeCoIn$_5$ we show that the large specific heat anomaly observed at $T_c$ at ambient pressure is suppressed
linearly at least up to 3 GPa.

\end{abstract}


\submitto{\JPCM}

\ead{gknebel@cea.fr}

\maketitle

\section{Introduction}
Heavy fermion systems provide the unique opportunity to study of the interplay of long range
magnetic order, unconventional superconductivity (SC) and valence fluctuations. For usual
superconductors the attractive interaction between two electrons forming a Cooper pair is due to a
lattice instability and magnetic impurities act as pair breaking. The finding of superconductivity
at the verge of an antiferromagnetic ordered state in cerium heavy fermion systems like
CeCu$_2$Si$_2$ \cite{Steglich79}, CeCu$_2$Ge$_2$ \cite{Jaccard92}, CeRh$_2$Si$_2$
\cite{Movshovich96} CePd$_2$Si$_2$ and CeIn$_3$\cite{Mathur98} suggested a pairing mechanism
associated with the magnetic instability. The importance of critical valence fluctuations for the
appearance of SC in systems with strong electronic correlations has been pointed out recently
\cite{Holmes04,Yuan03}.

The discovery of superconductivity in Ce$M$In$_5$ ($M$=Co, Rh, Ir) compounds opened new routes to
investigate the appearance of pressure induced SC in heavy fermion compounds and its interplay with
antiferromagnetism (AFM) \cite{Hegger00,Petrovic01a,Petrovic01b}. While CeCoIn$_5$ and CeIrIn$_5$
are superconductors at ambient pressure with superconducting transition temperatures $T_c=2.3$~K
and 0.4~K, CeRhIn$_5$ is antiferromagnetically ordered below the N\'eel temperature $T_N=3.8$~K and
SC appears only under hydrostatic pressure. The family of Ce$M$In$_5$ is closely related to
CeIn$_3$ and the crystal structure consists of alternating layers of CeIn$_3$ and $M$In$_2$
stacking along the $[001]$ direction. CeIn$_3$ is due to its cubic structure a nice model system to
study the appearance of superconductivity at a quantum critical point where the magnetic order is
suppressed, however SC appears only below 0.2~K in the pressure range of 2-3~GPa and the interplay
between AFM and SC is experimentally difficult to investigate \cite{Mathur98,Knebel01}. The
superconducting transition temperatures in CeCoIn$_5$ and CeRhIn$_5$ are enhanced by a factor of
almost 10 in comparison to CeIn$_3$. For superconductivity mediated by spin-fluctuations the higher
$T_c$ is expected for systems with lower dimensionality \cite{Monthoux01,Moriya03,Fukazawa03} and
indeed, in the 115 family the Fermi surface is almost two dimensional \cite{Shishido02}.

The pressure-temperature phase diagram of CeRhIn$_5$ has already been studied by resistivity $\rho$
\cite{Muramatsu01}, specific heat $C$ \cite{Fisher02}, magnetic susceptibility $\chi$, nuclear
quadrupole resonance (NQR) \cite{Kohori00,Mito01,Kawasaki01,Mito03,Kawasaki03}, and neutron
scattering experiments \cite{Bao01,Bao02,Majumdar02,Llobet04}. CeRhIn$_5$ orders at ambient
pressure in an incommensurate antiferromagnetic helical structure with a wave vector
$\bi{q}=(0.5,0.5,0.297)$ and a staggered moment of about 0.8~$\mu_B$. Contrary to the first
measurements \cite{Majumdar02}, recent neutron scattering measurements show no significant change
of the magnetic structure and the magnetic moment up to 1.7~GPa \cite{Llobet04}. However, a NQR
study shows that the internal magnetic field decreases linearly  with pressure and approaches
slowly a value of about 5\% at ambient pressure at 1.75~GPa \cite{Mito01,Kawasaki01,Mito03}. The
difference between neutron and NQR experiment is generally considered to be due to the different
time scale of the measurements. From all measurements, except the very first by Hegger {\it et
al.}, it follows that the antiferromagnetic order is suppressed near 2~GPa. Only the specific heat
experiments  \cite{Fisher02} found some anomaly above $T_c$ at 2.1 GPa: nevertheless, AFM order was
discarded as a possible origin for that anomaly  \cite{Fisher02}. SC has been found with transport
measurements in the pressure range from 1-8~GPa, with the maximum transition temperature
$T_c\approx 2.2$~K at $P\approx 2.5$~GPa \cite{Muramatsu01}. For pressures $P>2$ GPa CeRhIn$_5$
would be an unconventional superconductor with line nodes in the gap as shown by measurements of
the NQR relaxation rate $1/T_1$ which has a $T^3$ dependence below $T_c$  \cite{Kohori00,Mito01},
in agreement with specific heat measurements \cite{Fisher02}. In the intermediate pressure region
between 1.6 and  2 GPa, AFM and SC have been claimed to coexist, with possible  "extended gapless"
regions in the superconducting gap function. Recent NQR measurements claim to confirm this
possibility of gapless superconductivity in the coexistence regime from the observation of a
constant $T_1T$ below $T_c/2$, ascribed to a finite quasiparticle density of states
\cite{Kawasaki03}.

CeCoIn$_5$ is a unconventional SC with most probably $d$ wave symmetry and line nodes in the gap
\cite{Kohori00,Movshovich01,Izawa01,Ormeno02,Aoki03}.  At ambient pressure, it is located close to an
antiferromagnetic quantum-critical point (QCP). Detailed resistivity measurements show that
applying hydrostatic pressure tunes the system away from the proximity of the QCP
\cite{Nicklas01,Sidorov01}. The huge anomaly observed in specific heat at $T_c$ at ambient
pressure ($\Delta C /C(T_c) = 4.7$) decreases under pressure up to 1.5 GPa \cite{Sparn01}. De
Haas-van Alphen measurements show cyclotron masses at ambient pressure which are strongly field
dependent and decreasing under pressure \cite{Shishido03}.

In this article we report on detailed ac calorimetric measurements of CeRhIn$_5$ and CeCoIn$_5$ in
an extended pressure range up to 3.5~GPa. The measurements were performed in argon loaded diamond
anvil cells ensuring almost perfect hydrostatic pressure conditions. The main focus will be on the
appearance of SC in the coexistence phase of AFM and SC in CeRhIn$_5$. As the physical properties
of the 115 family are very sensitive to uniaxial pressure and pressure inhomogeneities
\cite{Oeschler03} hydrostaticity of the sample environment is very important. Previous specific
heat measurements on CeRhIn$_5$ were performed in piston cylinder type cell with a solid pressure
transmitting medium (AgCl) \cite{Fisher02}. Even if the pressure difference between different ends
of the sample is quite small, the effect of stress on the sample is not negligible. The nature of
the superconducting transition in CeRhIn$_5$ at high pressure will be related to that of
CeCoIn$_5$. The main result is for CeRhIn$_5$ the observation of nice specific heat anomalies at
the antiferromagnetic transition at low pressure and at the superconducting transition above 2~GPa.
Superconductivity appears in specific heat measurements only very close to the critical pressure
where both transitions are tiny and rather broad. From the pressure dependence of the
superconducting anomaly $\Delta C/C(T_c)$ it follows that in CeCoIn$_5$ at ambient pressure, SC
sets in when the effective mass of the electrons is still increasing towards low temperatures due
to the formation of the heavy fermion state, whereas at 3~GPa it behaves like a usual heavy fermion
superconductor. For both compounds the effect of pressure inhomogeneities on the magnetic and
superconducting transition will be discussed.

As regard notations, we call $P_{S-}$ the lowest pressure for which superconductivity is observed,
$P_{S+}$ the highest pressure for which superconductivity is observed, and $P_c$, the pressure of
the point where the AFM transition line $T_N(p)$ meets the superconducting transition
line $T_c(P)$. Let us remind here that the antiferromagnetic state is also labelled "AFM", the
superconducting state  "SC", a coexisting AFM and SC state  "AFM+SC", and the paramagnetic state
"PM".

\section{Experimental details}
High quality single crystals of CeRhIn$_5$ and CeCoIn$_5$ have been grown by the In flux method
\cite{Hegger00}. The specific heat measurements under pressure were performed
using ac calorimetry. In the case of CeRhIn$_5$ we set up two pressure cells
giving almost identical results. Details of this technique for measurements of the specific heat
are given elsewhere \cite{Sullivan68,Demuer00,Wilhelm02}. The samples studied were of size of about
$200 \times 200 \times 60\mu$m. An AuFe/Au thermocouple served to measure the temperature
oscillations of the sample. It is soldered directly on the sample to ensure a good thermal
contact between thermometer and sample. As heater we used a 50 mW argon laser. By using a
mechanical chopper it is possible to obtain a quasi sinusoidal power which is transmitted by
optical fibre directly to the sample. However, this method doesn't allow quantitative measurements,
as the heating power is not focused on the sample, but irradiates also the pressure transmitting medium
(argon) and the gasket which are heated and contribute to an additional background signal which changes
between different experiments. To find the optimal working frequency $\nu$, the frequency dependence
of the ac signal was measured at 1.5~K and 4.2~K. The cut off frequency $\nu_c$  was found to be
about 600~Hz, the measurements were performed at 831~Hz slightly above $\nu_c$. The specific heat
of the sample can be estimated by $C_{ac} \propto -P S_{th} \sin (\theta -\theta_0) / V_{th} 2\pi
\nu$ where $V_{th}$ and $S_{th}$ are respectively the measured voltage and the thermopower of the
thermocouple. As the origin of the phase $\theta_0$ cannot be determined by our method, we neglect
in the analysis the contribution of the phase signal. However a comparison of the behaviour of the
signal at low pressure with an absolute measurement at ambient pressure shows, that observed ac
signal is correct.

The ac susceptibility was measured in an argon loaded sapphire anvil cell with 2.5~mm tables diameter. Both
anvils are placed inside one of the detection coils (5000 turns), the second detection coil is
placed above the anvils. In this geometry the sample and the gasket are in the middle of the lower
detection coil. This geometry allows a very good compensation of the susceptometer at fixed
temperature, however the filling factor is poor. An additional difficulty is coming from a temperature
drift of the background signal which cannot be compensated. The measurements were performed at
71~Hz, and before each run the susceptometer was offset at the lowest temperature by compensating
the amplitude and the phase of the signal with a small compensation coil which is wounded directly
on the excitation coil. This susceptometer allows the detection of the onset temperature
of the superconducting transition due to the diamagnetic shielding, however it is not possible to conclude
about the superconducting volume fraction. The total volume of the measured samples was about
0.01~mm$^3$.

In both experiments the pressure was determined in-situ at low temperatures by the  ruby fluorescence
at 4.2~K. A bellow system allows to change and fine tune the pressure at low temperature
\cite{Salce00}.

\section{Specific heat of CeRhIn$_5$ under pressure}
\subsection{Experimental results}

\begin{figure}
\begin{center}
\scalebox{0.6}{\epsfbox{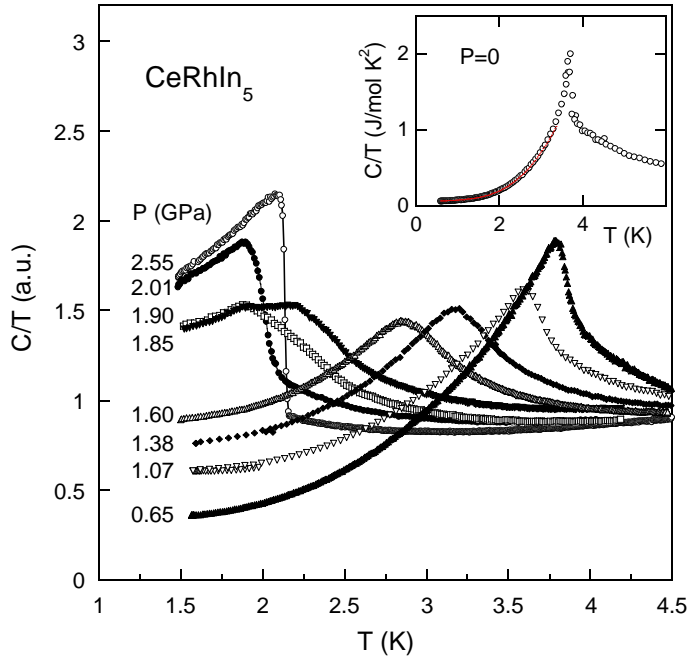}}
\end{center}
\caption{\label{figure:CeRhIn5_specific_heat1}Specific heat of CeRhIn$5$ for different pressures
$p$ (pressure cell \#1). The data are normalized at $T=5$~K. The inset shows the specific heat
measured at ambient pressure.}
\end{figure}

\begin{figure}
\begin{center}
\scalebox{0.6}{\epsfbox{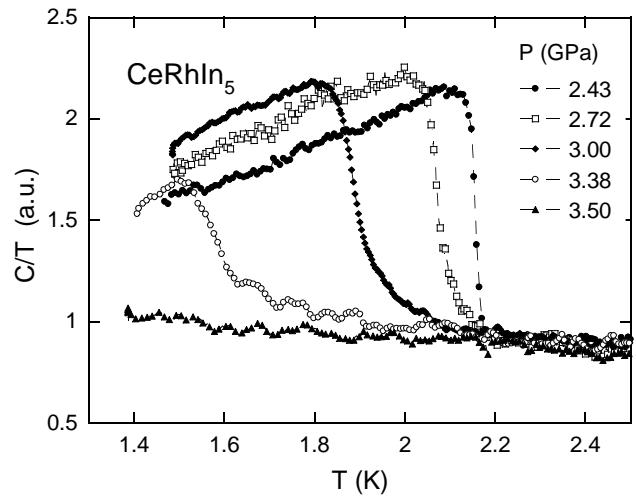}}
\end{center}
\caption{\label{figure:CeRhIn5_C(T)_supra}Superconducting transition at high pressures for
CeRhIn$_5$ (pressure cell \#2). $C/T$ is normalized in the nomal state at $T=2.2$~K.}
\end{figure}

The temperature dependence of the specific heat signal of CeRhIn$_5$ is plotted in figure
\ref{figure:CeRhIn5_specific_heat1} for different pressures. The inset shows the specific heat of
CeRhIn$_5$ at ambient pressure. At $T_N$, $C/T$ has a very sharp peak at ambient pressure. The
entropy connected with the magnetic transition is small, of about $0.3R\ln 2$. The remaining
entropy is recovered up to 20 K. This strong enhancement of the specific heat in the vicinity of
$T_N$ shows the importance of short range order (magnetic fluctuations) and is not described by
mean field theory. In the magnetically ordered state, $C/T$ can be best approximated by taking into
account an electronic contribution $C_{el}/T =\gamma + \beta_M T^2$ and an additional term
corresponding to an antiferromagnetic spin wave with a gap in the excitation spectrum $C_{g}/T =
\beta'_M e^{-\Delta/T}$ \cite{Cornelius00}. As parameter we find $\gamma = 52$ mJ/mol K$^2$, $\beta_M = 24$ mJ/mol K$^4$, $\beta_M = 756$ mJ/mol K$^4$, and $\Delta = 8.1$ K. In comparison to the anomaly at $T_N$ at ambient pressure, at 0.6 GPa the magnetic anomaly is shifted to higher temperatures and the transition is
only slightly broadened. The magnetic ordering temperature $T_N$ is determined by the maximum of
$C/T$. With increasing pressure above 0.6~GPa, $T_N$ is decreasing and for pressures higher than 1
GPa the transition starts to broaden, however the magnetic anomaly remains well defined. At 1.85
GPa the magnetic anomaly at $T_N = 2.2$~K is very broad. A second maximum associated with a
superconducting transition is observed at lower temperatures at $T_c=1.8$~K. Increasing the
pressure by only 0.05 GPa leads to a suppression of the maximum at the magnetic transition, and
only a shoulder above $T_c$ points to an antiferromagnetic state. With further increasing pressure,
the superconducting transition gets more pronounced and at 2 GPa, slightly above $P_c=1.95$~GPa,
only a clear superconducting transition is found. In the investigated temperature range $T>1.4$~K
we see no sign of a magnetic transition above $P_c$ in the superconducting phase. The
superconducting transition increases up to 2.21~K at 2.4 GPa. Increasing further the pressure leads
to a suppression of $T_c$ (see figure \ref{figure:CeRhIn5_C(T)_supra}).

\begin{figure}
\begin{center}
\scalebox{0.6}{\epsfbox{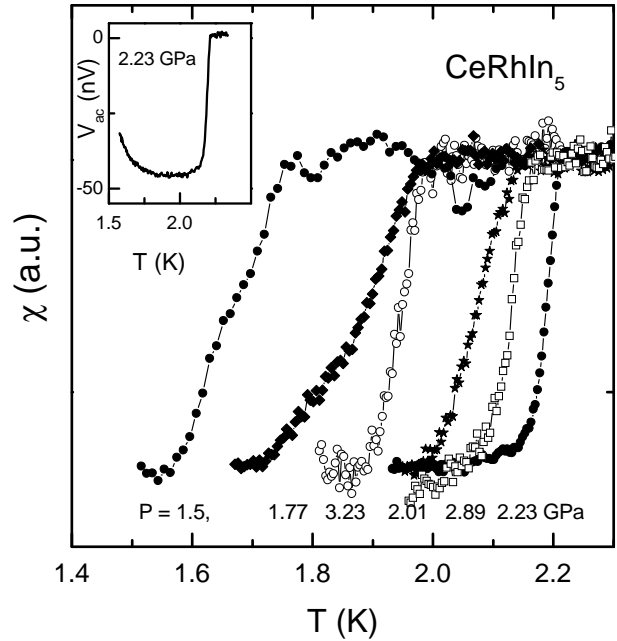}}
\end{center}
\caption{\label{figure:CeRhIn5_suscept}Superconducting transition of CeRhIn$5$ observed in ac
susceptibility. The amplitude of the transition is arbitrary and scaled for different pressures.
The onset of the transition is very sharp only above the critical pressure $P_c=1.9$ kbar. The
inset shows the observed signal in absolute units for 2.23 GPa. The increase of the signal to low
temperatures is due to the background. }
\end{figure}

In figure \ref{figure:CeRhIn5_suscept} the ac susceptibility signal connected with the
superconducting transition is plotted. A superconducting anomaly is first seen at 1.5~GPa, but the
width of the transition $\Delta T_c = 200$~mK is very large. With increasing pressure the
transition width gets smaller and $T_c$ is increasing ($\Delta T_c = 50$~mK at 2.3~GPa) where the
maximum $T_c = 2.21$~K is observed. For $P<P_c$ and for $P>2.5$~GPa the onset of the
superconducting transition by susceptibility ($T_c^\chi$) is at higher temperatures than the onset
of the transition by specific heat ($T_c^C$). A cascade $T_c^\rho>T_c^\chi>T_c^C$ of
superconducting transition temperatures determined by resistivity $(T_c^\rho)$, susceptibility and
specific heat measurements is characteristic of heterogeneous material.

\subsection{Phase diagram of CeRhIn$_5$}

\begin{figure}
\begin{center}
\scalebox{0.6}{\epsfbox{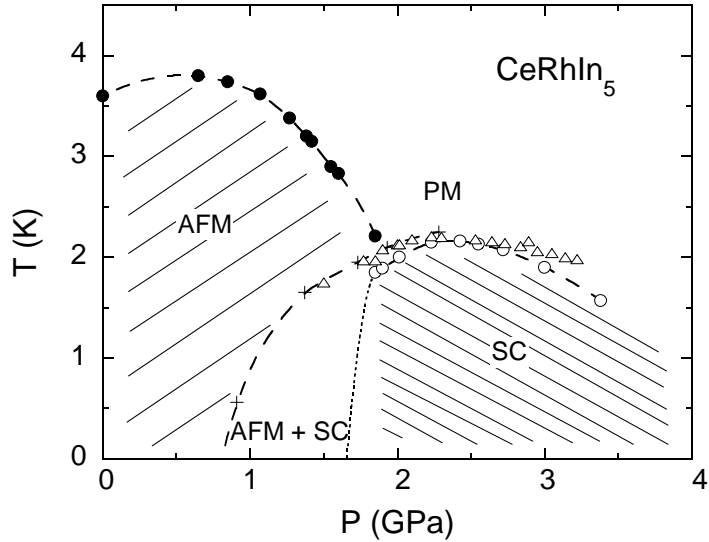}}
\end{center}
\caption{\label{figure:CeRhIn5_phase_diagram}Phase diagram of CeRhIn$_5$ as determined by specific
heat  ($\bullet$ and $\circ$) and susceptibility measurements ($\bigtriangleup$). In addition $T_c (p)$ from resistivity measurements ($\rho = 0$) after Llobet {\it et
al.}\cite{Llobet04} is plotted (+). The hatched areas mark the regimes where pure AFM and pure SC states are observed. The dotted line indicates the most probable first order line between the AFM and the SC bulk phase; in the (AFM+SC) regime SC is expected to be only filamentary and not a bulk property. }
\end{figure}

In figure \ref{figure:CeRhIn5_phase_diagram}, we summarize the phase diagram of CeRhIn$_5$ obtained
by specific heat and susceptibility measurements. In addition we plotted $T_c$ obtained by
resistivity measurements $(+)$ from Llobet {\it et al.} \cite{Llobet04}. The phase diagram of
CeRhIn$_5$ can be divided in three different parts: at low pressure, $P<0.9$ GPa, the ground state
is purely antiferromagnetic. In a limited pressure range 0.9 GPa$ < P < 1.95$~GPa superconductivity
and antiferromagnetism may coexist, and for $P>1.95$~GPa the ground state is superconducting. The AFM transition line $T_N(P)$ meets the SC transition line at $P_c \approx 1.95$~GPa. 

Let us first discuss the AFM transition. At low pressures $T_N (P)$ is first increasing with pressure and has a smooth maximum at $P\approx 0.6$~GPa. In the intermediate pressure range 0.9~GPa$<P<$ 2~GPa,
$T_N$ is monotonously decreasing up to 1.9~GPa, with a continuously increasing rate exceeding
2~K/GPa at $P_c$. Near $P_c$, the magnetic transition gets very broad and the amplitude of the
magnetic transition is strongly decreasing compared to the low pressure measurements. These
broadening effects will be more quantitatively discussed in the next section. Let us note that both
NQR and specific heat measurements agree on the fact that above $P_c$, no AFM order is observed :
the ground state for $P>P_c$ {\it is a pure superconducting state}. As regard even the latest
neutron measurements \cite{Llobet04}, they do not extend beyond $1.85$~GPa. However, the strange
result (in apparent contradiction with the slow NQR probe ($10^{-7}$~s)) is that the low temperature
ordered moment determined by a quasi-instant probe as neutron scattering ($10^{-11}$~s) does not
collapse with $T_N$ close to $P_c$, but the staggered moment is almost constant up to 1.85 GPa.

Switching now to the superconducting transition, the most remarkable fact is the absence of AFM
order below $T_c$ above $P_c$: for $P>P_c$ the ground state is a pure superconducting state. Nice
superconducting anomalies are observed, becoming sharper near the maximum $T_c=2.2$~K at 2.55~GPa.
At this pressure, the transition width is comparable to the superconducting transition in
CeCoIn$_5$ at ambient pressure. For higher pressures $T_c$ determined from the specific heat
experiment decreases with a rate of -0.7K/GPa. Resistivity measurements by Muramatsu {\it et al.}
show that superconductivity is completely suppressed at a pressure $P_{S+}$ of $\approx 8$~GPa
\cite{Muramatsu01}. Contrary to the previous work \cite{Fisher02}, we do not observe any round anomaly
above $P_c$ due to its normale phase, and a fortiori no sign of AFM transition. In agreement with \cite{Fisher02}, we also
do not observe any sign of an AFM transition below $T_c$, even very close from $P_c$. So in
CeRhIn$_5$, $T_N$ is not suppressed continuously to zero, but has a finite value at $P_c$. It
demonstrates the absence of a quantum critical point in CeRhIn$_5$. Thermodynamically,  it means
that once $T_c$ is above $T_N$, the free energy of the superconducting state is lower than that of
the AFM state, whatever the temperature, and that in CeRhIn$_5$, contrary to the usual consensus on
heavy fermion superconductors, AFM order and superconductivity compete.

This has also consequences on the pairing mechanism: this competition and the closeness of the
energy scales of both phenomenon, makes the AFM correlations as a sole source of the pairing
mechanism very unlikely. For example, an extraordinary strong coupling regime would be required to
explain that the maximum $T_c$ is so close (a factor of $\approx 2$) to the maximum $T_N$. Further, the superconducting anomaly $\Delta C/C(T_c)$ is largest in the pressure range from 2.5~GPa to 3~GPa, pointing to the maximum of the pairing interation for SC above $P_c$. Interestingly this is the pressure range where the resistivity has a linear temperature dependence above $T_c$ and the residual resistivity $\rho_0$ a maximum as function of pressure \cite{Muramatsu01}. This points to the probable importance of valence fluctuations in the superconducting pairing mechanism \cite{Holmes04}. 

The regime above $P_c$ puts also severe constraints on the possible coexistence regime below $P_c$.
In resistivity measurements on high quality crystals in Los Alamos\cite{Llobet04}, SC is
found down to much lower pressures than $P_c$: $P_{S-}\approx~0.9$ GPa. The
extrapolation of $T_c \to 0$ coincides almost with the pressure of the maximum of $T_N$. The
transition temperatures $T_c$ determined from the ac susceptibility measurements are in good
agreement with these resistivity measurements. However, with ac calorimetry, we  find a
superconducting anomaly only very close to the critical pressure $P_c=1.95$~GPa. This questions the
homogenous coexistence of superconductivity and antiferromagnetism in this pressure range, as the
observed transitions in resistivity and susceptibility are not a bulk probe of superconductivity.
From the specific heat measurements, we know that at 1.5 GPa, $T_c$, if non zero, is below 1.5 K.
So instead of $P_{S-}\approx~0.9$ GPa, we expect an almost vertical line between $P_{S-}$ and
$P_c$. This would mean that the line $T_c(P)$ drawn by resistivity or susceptibility measurements
within the AFM state does not reflect a bulk transition, and might be connected to internal stress
inside the sample, like in CeIrIn$_5$ \cite{Bianchi01}. This also means that previous claims of a
coexistence of AFM order and superconductivity \cite{Llobet04, Kawasaki03} relying on the
observation of AFM order below the resistive $T_c$ in the pressure range between
$P_{S-}$ and $P_c$ are not a definite proof of that coexistence. Differences in $T_c^\chi$ and $T_c^C$ are also observed above 2.5~GPa, the pressure of maximum $T_c$. Of course, in our scenario of a direct AFM $\to$ SC transition, the line between AFM and SC is expected to be a first order line, owing to the
sudden disappearance of the magnetic order parameter (and in agreement with the strong slope of
$\vert\frac{\partial T_c}{\partial P}\vert$). Further intrinsic phase separation with a mixed phase may be possible. In CeIn$_3$, phase separation was nicely demonstrated by NQR \cite{Kawasaki04}.

To summarize this discussion, from our specific heat and ac susceptibility measurements, two
different scenarii are possible; (i) the appearance of superconductivity in the antiferromagnetic
ordered state is not homogenous and no true AFM+SC state exists. The experimental observations
would then result from superconducting filaments, which can be created due to internal stress
induced by dislocations or stacking faults, or due to a phase segregation in a pure magnetically
ordered and in a superconducting volume fraction. With increasing pressure the antiferromagnetic
volume is decreasing and the paramagnetic volume which has a superconducting ground state
increases. Above $P_c$, only the superconducting state survives. (ii) The coexistence is really
homogenous, which means that the same electrons are responsible for the antiferromagnetic order and
for superconductivity. In this case the missing anomaly of specific heat is due to a gapless
superconducting state which is not explained by impurities, and the coexistence phase corresponds
to a new class of superconducting states \cite{Fuseya03}. In the following, both possibilities will
be discussed, also we strongly believe in the first scenario.

\subsection{On the transition broadening}

\begin{figure}
\begin{center}
\scalebox{0.6}{\epsfbox{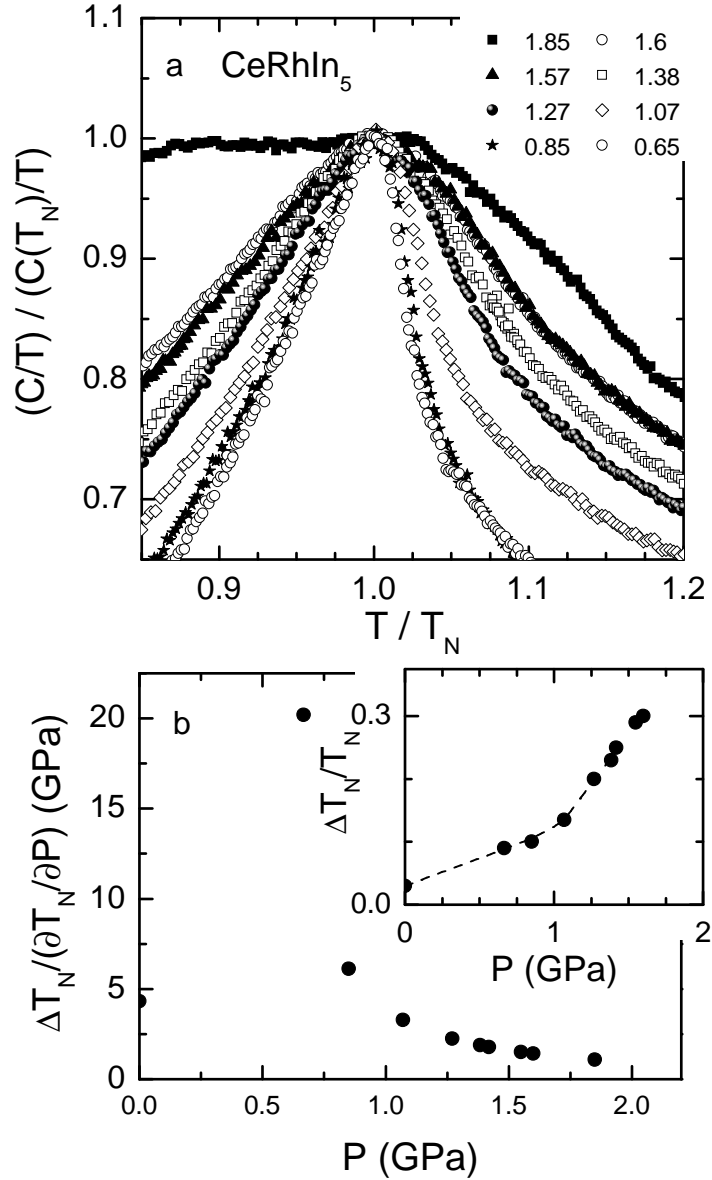}}
\end{center}
\caption{\label{figure:CeRhIn5_boadening_afm}Upper frame a) Specific heat anomaly  due to the
antiferromagnetic order of CeRhIn$5$ normalized by the maximum of $C/T$ at $T_N$ as function of
$T/T_N$. Lower frame b) the transition width $\Delta T_N$ normalized by the slope of the variation
of $T_N$ versus $p$.  The width of the transition is arbitrarily taken at $(C(T)/T)/(C(T_N)/T)=0.8$. The inset shows the relative width of the antiferromagnetic  transition  $\Delta T_N /T_N$
observed in the specific heat experiment as function of pressure. }
\end{figure}

Figure \ref{figure:CeRhIn5_boadening_afm}a shows the specific heat in a normalized representation
$\frac{C(T)/T}{C(T_N)/T}$ as function of $T/T_N$. To quantify the observed broadening of the
magnetic anomaly, we arbitrarily define the full width of the transition when
$\frac{C(T)/T}{C(T_N)/T} =0.8$. The relative width of the antiferromagnetic transition as a
function of pressure is then shown in the inset of figure \ref{figure:CeRhIn5_boadening_afm}b. A
change of regime is clearly visible at 0.9~GPa where $T_N$ starts to decrease and superconductivity
is observed by Llobet et al.\cite{Llobet04}. At low pressure the anomaly is rather sharp, but for
$P>0.9$~GPa it gets continuously broader. However, several effects come into play. Part of the
width is intrinsic, coming from the fluctuations, and it is expected to yield a constant value of
$\frac{\Delta T_N}{T_N}$. In addition, material inhomogeneities, internal or external stress,
pressure gradients, might give a pressure dependent contribution to $\left(\frac{\Delta
T_N}{T_N}\right)$. Some of these effects will be proportional to the pressure variation of $T_N$and
so to $\frac{\partial T_N}{\partial P}$ in a first approximation (see
\ref{figure:CeRhIn5_boadening_afm}b). Detailed measurements show that the pressure variation in a
diamond anvil cell with argon in the low pressure range ($P<6$~GPa) is generally lower than
0.04~GPa \cite{Thomasson04}. Considering the width and also the shape of the ruby spectra, we could
not detect any significant broadening of these spectra over the whole investigated pressure range.
If the observed broadening would result only from pressure inhomogeneities in the pressure cell,
this would require pressure inhomogeneities of the order of 0.055~GPa near $P_c = 1.95$ GPa, which
can be excluded.

Further, the observed very sharp superconducting transition at 2.4 and 2.55~GPa in different
pressure cells are a posteriori a strong indication of high hydrostaticity. Here the width of the
transition, $\Delta T_c = 3$~mK, is comparable to the superconducting transition of CeCoIn$_5$ at
ambient pressure (see below). This clearly shows that the broadening of the magnetic transition on
approaching the critical point is not related to the pressure cell. Similar behaviour has been
observed in other heavy fermion systems like CeIn$_3$ \cite{Knebel01}, CePd$_2$Si$_2$
\cite{Demuer01} and CeRh$_2$Si$_2$ \cite{Haga04} too. Theoretically, in the case of a second order
phase transition which ends at the critical pressure $P_c$, the form of the mean field magnetic
transition for $T_N \to 0$ is sharp. Only the size of the anomaly is decreasing as the ordered
magnetic moment is decreasing \cite{Zuelicke95}, which is not even the case in CeRhIn$_5$
\cite{Llobet04}. As regard impurity effects, in the classical framework of a second order phase
transition (Harris criterion \cite{Dotsenko95}), they are believed to change the critical behaviour only if the specific heat diverges at $T_N$.

Here the phenomena is quite different and more similar to surface problems
found in magnetism or for some local structural transitions. Physically it seems that the magnetic coherence length at $T_N$ cannot exceed a critical value $\xi_c$. As for $P\to P_c$ the magnetic coherence length at $T \to 0$ will increase strongly, there is a severe cut off in the development of a large coherence length and thus in a corresponding smearing of the specific heat anomaly.  

In the pressure range of a first order transition the entropy drop $\Delta S$ associated to the
magnetic transition is linked to the slope of $\partial T/\partial P$ according to the Clapeyron relation  $\partial T/\partial P = \Delta V/\Delta S$, where is $\Delta V$ is the volume 
discontinuity. The final vertical slope of $\partial T/\partial P$ as $T \to 0$ reflects the
collapse of the entropy in agreement with the Nernst principle. As the material is highly sensitive
to imperfections, the entropy drop corresponds to a broad specific heat anomaly. The corresponding
entropy contribution $\Delta S$ reflects the amplitude of $\partial T/\partial P$ as by contrast
the volume discontinuity may change weakly under pressure. So it is quite reasonable that the
specific heat anomaly of the magnetic transition disappears drastically on approaching $P_c$.

\begin{figure}
\begin{center}
\scalebox{0.6}{\epsfbox{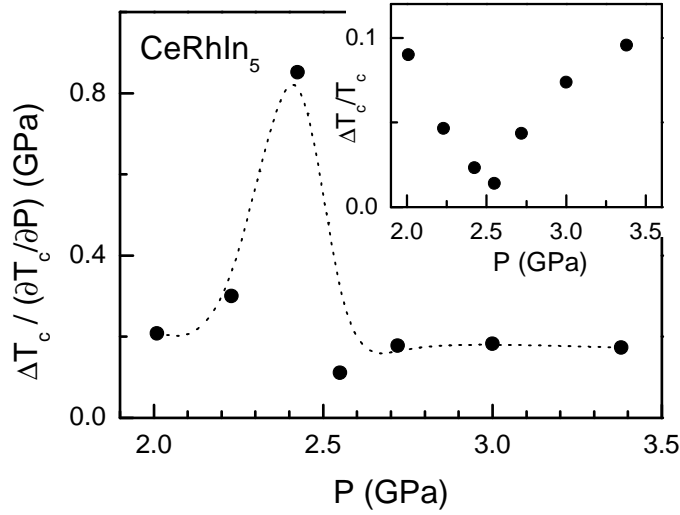}}
\end{center}
\caption{\label{figure:CeRhIn5_boadening_supra}Width of the superconducting transition in
CeRhIn$_5$ in the ac calorimetry normalized by the slope of $T_c$ versus $P$. It is basically
constant, the high point at $P_c$ coming from the residual finite width divided by a vanishing
$\frac{\partial T_c}{\partial P}$. Inset: Relative width of the superconducting transition from
specific heat experiments in CeRhIn$_5$ as function of pressure.}
\end{figure}

Internal stress may lead to drastic features as antagonistic behavior are
often observed in the variation of the N\'eel temperature for a strain $\sigma$ applied in nonequivalent directions. Well known examples for tetragonal systems are
 CePd$_2$Si$_2$ \cite{vanDijk00,Demuer02} or URu$_2$Si$_2$
\cite{deVisser86,vanDijk95}. In the last one the values at ambient pressure are: $\partial
T_N/\partial \sigma_a = +900$~mK/GPa and $\partial T_N/\partial \sigma_c = -410$~mK/GPa for $T_N =
17$~K. The strain dependence of
$T_c$ in URu$_2$Si$_2$ illustrates the antagonism between magnetism and superconductivity as the
respective variations of $T_c$ and $T_N (\sigma)$ are opposite: $\partial T_c/\partial \sigma_a =
-620$~mK/GPa and $\partial T_c/\partial \sigma_c = +430$~mK/GPa for $T_c = 1.2$~K. Huge effects have also been detected in the pressure dependence of $T_N$ in CePd$_2$Si$_2$
measured on two crystals with the $c$ axis parallel or perpendicular to the pressure gradient due
to the non-hydrostaticity of the pressure cell \cite{Demuer02}. Of course, as the superconducting
domain is locked at $P_c$, a large difference appears also in $T_c (P)$. To summarize the
broadening is strongly associated to the magnitude of $\partial T_N/\partial P$ and thus the
calorimetric anomaly collapses at $P_c$.

In figure \ref{figure:CeRhIn5_boadening_supra} the transition width of the superconducting
transition above $P_c$ is shown as function of pressure, and in figure
\ref{figure:CeRhIn5_boadening_afm} that of the AFM transition below $P_c$. In this system, due to
the competition between AFM and SC order, we rather expect that these respective transition width
are related to the strength of the pressure variation of their own critical temperature. What is
more, in the 115 series, the main source of heterogeneities  may be that internal pressure or
strain gradients in the material itself. Thus, local distributions of $T_c$ or $T_N$ may be
induced. It is well known that near defects like dislocations or stacking faults, internal strain
of the order of 0.1 GPa can occur. Evidences for such an effect are given in the paramagnetic state
of CeIrIn$_5$ at zero pressure as  $T_c^\rho = 1.2$ K is quite different from $T_c^C = 0.4$ K
\cite{Bianchi01}. In that case, the superconducting transition observed by resistivity is clearly
due to superconducting filaments. This big mismatch of $T_c$ as measured by resistivity or specific
heat seems again directly linked with the difference between $\partial T_c/\partial \sigma_a
=540$~mK/GPa and $\partial T_c/\partial \sigma_c =-840$~mK/GPa \cite{Oeschler03}. Of course an
extra cause of heterogeneity can be induced by the non hydrostaticity of the pressure transmitting
medium. However, in our case, the use of argon optimizes the hydrostaticity.

\begin{figure}
\begin{center}
\scalebox{0.6}{\epsfbox{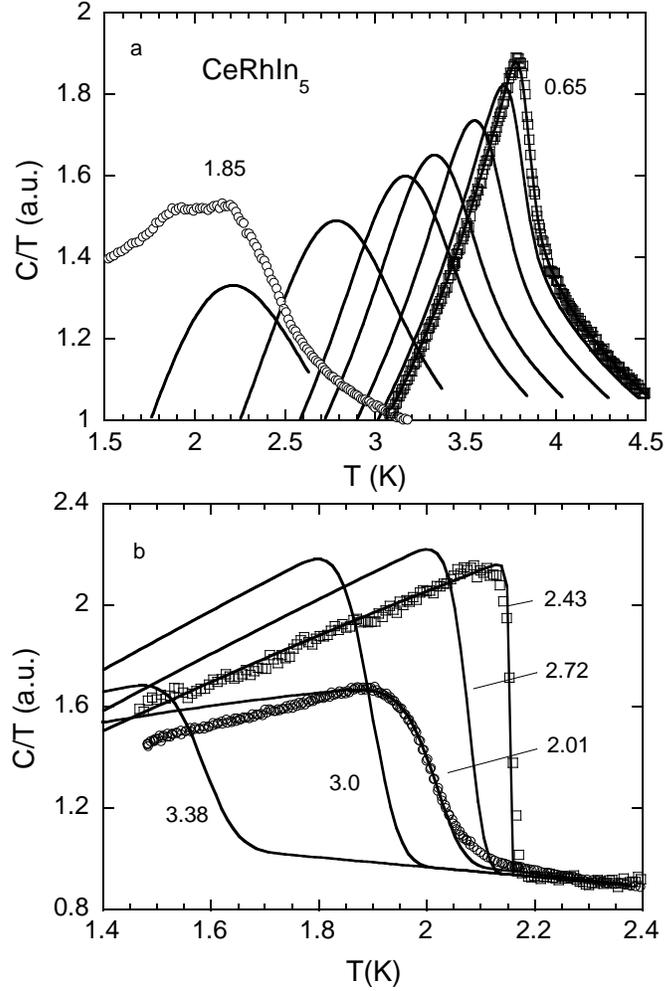}}
\end{center}
\caption{\label{figure:CeRhIn5_cp_fit}Upper frame a: Modelling of the specific heat of CeRhIn$_5$
taking into account a pressure distribution of $\Delta P = 0.045$~GPa and the slope of $\partial
T_N/\partial p$ for different pressures ($P$=0.65, 0.85, 1.07, 12.7, 13.8, 1.6 and 1.85 GPa). For
comparison the measured specific heat for $P=0.65$ GPa and 1.85 GPa is plotted. Lower frame b:
Effect of the same pressure distribution of $\Delta P = 0.055$~GPa on the superconducting
transition.}
\end{figure}

To demonstrate the impact of internal strain or pressure inhomogeneities when $\partial
T_N/\partial P$ and $\partial T_c/\partial P$  have a strong pressure dependence, we have
calculated the temperature dependence of the specific heat near the antiferromagnetic and the
superconducting transition under the assumption of a pressure distribution inside the sample, of
width $\Delta P$, which may be caused by the experimental conditions or by inhomogeneities in the
material. For the antiferromagnetic transition (see figure \ref{figure:CeRhIn5_cp_fit}a) we suppose
that in an hypothetical ideally hydrostatic pressure cell, the shape of the specific heat anomaly
would remain unchanged whatever the transition temperature. We have further assumed that at low
pressure, the pressure variation of $T_N$ is small and we can take the curve at 0.65 GPa as the
ideal curve. Indeed, $\partial T_N/\partial P \approx 0$ for 0.65~GPa, so that pressure gradients
should have only minor effects on the shape of $C/T$. We assume a gaussian pressure distribution
inside the sample, so that the form of the specific heat anomaly for an mean average pressure $P_0$
is given by
\[
\left.\frac{C}{T}\right|_{P_0}(T) = \int_{}^{}1/W(P)
\exp\left[-\frac{1}{2}\left(\frac{P-P_0}{\Delta P}\right)^2\right]
\left.\frac{C}{T}\right|_{0.65}\left(\frac{T}{T_N(P)}\right) dP.
\]
The weighting factor W includes the normalization of the gaussian distribution, and normalization
of $\left.\frac{C}{T}\right|_{0.65}\left(\frac{T}{T_N(P)}\right)$ with respect to entropy balance:
$W(P)$ should be proportional to $T_N$ for localized magnetism, or constant for itinerant
magnetism. The difference in the resulting curves is found to be insignificant for the pressure
distribution involved in this experiment. Calculated specific heat transitions for the experimental
pressures $P_0$ and a pressure distribution of $\Delta P = 0.055$~GPa are shown in figure
\ref{figure:CeRhIn5_cp_fit}. They have to be compared to the measurements (see figure
\ref{figure:CeRhIn5_specific_heat1}). The broadening in the range where $\partial T_N/\partial P$
is steep, is clearly visible in the calculations and it is in qualitative agreement with the
measurements.

A more quantitative comparison between experiment and these calculations has been done for the
superconducting transition, with the same pressure distribution $\Delta P = 0.055$~GPa. We have
calculated the specific heat near the superconducting transition for $P>2$~GPa (see figure
\ref{figure:CeRhIn5_cp_fit}b). The entropy balance imposes that $S_n = S_s$ at $T_c$ for all
pressures. We have assumed a power law dependence of $C/T (P) = A (T/T_c)^\alpha$ in the
superconducting state \cite{footnote}, adjusting the exponent $\alpha$ (which depends on the
relative specific heat jump at $T_c$) in order to fulfill the entropy balance. So calculation of
$C/T$ for a fixed pressure distribution $\Delta P$ is controlled by two parameters: $T_c(P)$, which
is known from the phase diagram, and the size of the anomaly at the average pressure (which
controls $\alpha$). The comparison with the measurements (see figure
\ref{figure:CeRhIn5_C(T)_supra}) shows that the broadening of the transition for pressures below
and above the maximum of $T_c (P)$ can be understood with the same fixed pressure distribution.

To summarize, a fixed gaussian pressure distribution of about 0.055~GPa can explain the observed
broadening of the antiferromagnetic and the superconducting transitions, as a result of the
pressure dependence of $T_N$ and $T_c$. As regard the origin of this pressure distribution,
0.04~GPa is really the upper limit expected for inhomogeneities inside a pressure cell filled with
argon. More reasonably, these inhomogeneities could be due to internal strain and defects in
combination with the anisotropic elastic properties of the material \cite{Kumar04}.

\subsection{On the possibility of gapless nature of superconductivity: material effects or novel phase ?}
The question of a gapless SC state below $P_c$ started with recent NQR results \cite{Kawasaki03}:
just above $P_c$, the nuclear relaxation time follows the usual behavior of an unconventional SC
state with line nodes: below $T_c$, $(T_1T)^{-1}$ has a nice $T^2$ dependence. This contrasts with
the situation below $P_c$. For example at 1.6 GPa, below $T_c^\chi$, $(T_1T)^{-1}$ reaches rapidly
at low temperatures $(T \ll T_N)$
 a value corresponding to the normal phase \cite{Kawasaki03}. According to Fisher {\it et al.} \cite{Fisher02}, the specific
heat coefficient $\gamma$  increases by a factor 3 to 4 from 0 to 1.6~GPa. This increase of the
effective mass leads to an increase of  $(T_1T)^{-1}$ by one order of magnitude, as observed in the
experiment.

We have stressed that our measurements and analysis do not support an intrinsic AFM+SC state
between $P_{S-}$ and $P_c$. However, a gapless state in this pressure region could be possible
without any additional line in the phase diagram (a continuous evolution of the gap amplitude
collapsing when on approaching $P_c$ from the high pressure region would not necessarily induce
symmetry changes). This gapless state cannot be due to impurity scattering, as the criterion for
clean limit are satisfied as well below and above $P_c$: the sample investigated in our measurement
has a residual resistivity ratio of almost 200 which shows that it is very clean.

The possibility of the realization of $p$-wave spin singlet superconductivity, whose gap function
is odd in frequency and momentum, was very recently discussed by Fuseya et al.\cite{Fuseya03}. They
showed that near a quantum critical point where strong retardation effects are possible, this
$p$-wave state is more likely than the $d$-wave state which is expected to be realized away from
the critical point in the antiferromagnetic as well as in the paramagnetic regime. A quantum
critical point is not observed in our experiment, and a gapless region is also not observed above
$P_c$ \cite{Fisher02}. Nevertheless, this difference might arise from the first order nature of the
AFM $\to$ SC transition. The NQR results were interpreted with a heuristic view in favour of this
new class of superconducting phase below $P_c$ which differs from the usual $d$ wave pairing
\cite{Kawasaki03}. Basically, the bare experimental features are similar to those observed here:
$T_c^\rho$, the superconducting onset chosen in resistivity is higher than $T_c^\chi$ where
diamagnetic shielding is observed.  $T_c^\chi$  appears to coincide with the temperature where tiny
features appear in the temperature variation of $(T_1T)^{-1}$ of the inverse product of the nuclear
relaxation time $T_1$ by temperature.

From our point of view, the difficulty with this scenario is both quantitative: it is not expected
that $T_c$ could rise up to $T_N$ (at $P_c$), and qualitative: switching from a gapless $p$-wave
state below $P_c$ to a gaped $d$-wave state above $P_c$ would involve a symmetry change and thus
transform the tricritical point at $P_c$ to a tetracritical point. We would rather interpret the
"gapless" nature of superconductivity observed by NQR as related to the heterogeneities observed in
the magnetic transition. However, it is obvious that the debate remains. An important issue is to discuss more deeply the discrepancy between neutron diffraction and NQR measurements in the AFM phase.

\section{Superconductivity in CeCoIn$_5$ under high pressure}

\begin{figure}
\begin{center}
\scalebox{0.6}{\epsfbox{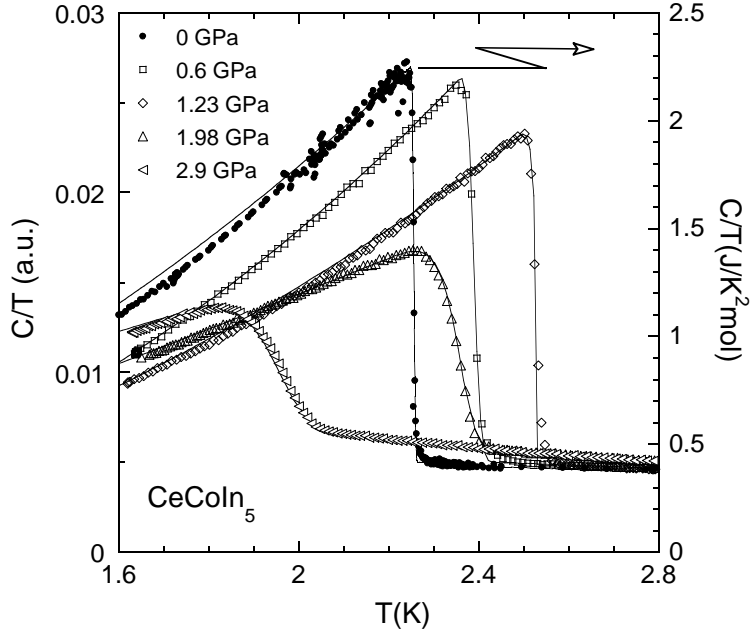}}
\end{center}
\caption{\label{figure:Cac_CeCoIn5_fit}Specific heat of CeCoIn$_5$ under high pressure for
different pressures. The ac signal is corrected by a constant background of about 40\% of the
measured specific heat in the normal state and it is assumed to be linear in temperature and
independent of pressure. For comparison, $C/T$ determined by a quantitative measurement is shown
($P=0$, right scale). Lines are calculations of the specific heat under the assumption of a
pressure distribution $\Delta P$ which increases linearly under pressure from $\Delta P=0.015$ for
$P=0$ up to $\Delta P=0.15$ for $P=2.9$~GPa. }
\end{figure}

\begin{figure}
\begin{center}
\scalebox{0.6}{\epsfbox{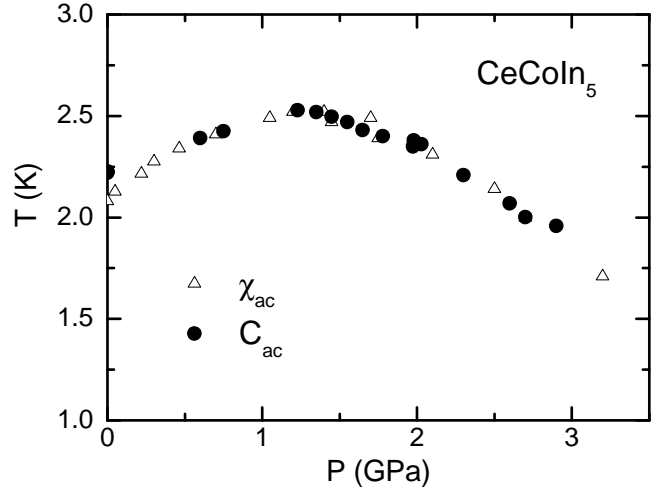}}
\end{center}
\caption{\label{figure:CeCoIn5_phase_diagram}Pressure-temperature phase diagram of CeCoIn$_5$
obtained by high pressure ac calorimetry ($\bullet$) and ac susceptibility ($\bigtriangleup$).}
\end{figure}

\begin{figure}
\begin{center}
\scalebox{0.6}{\epsfbox{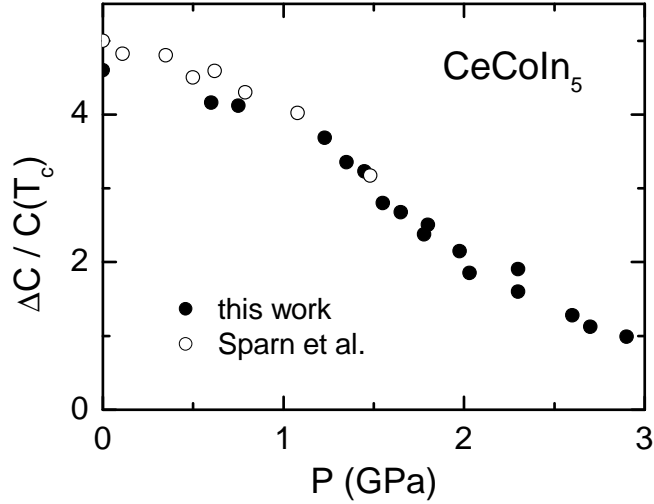}}
\end{center}
\caption{\label{figure:CeCoIn5_delta_C}Pressure dependence of the jump of the specific heat at
$T_c$ in CeCoIn$_5$. ($\circ$) are taken from reference \cite{Sparn01} and are used to normalize
the jump in the ac calorimetry.}
\end{figure}

The specific heat under high pressure of CeCoIn$_5$ is shown in figure \ref{figure:Cac_CeCoIn5_fit}.
Up to 1.5~GPa, the anomaly under high pressures is almost as sharp as at ambient pressure, whereas at higher
pressures the anomaly start to get broader. The phase diagram obtained from specific heat and ac
susceptibility measurements is shown in  figure \ref{figure:CeCoIn5_phase_diagram}. With increasing
pressure $T_c$ is increasing with an initial rate of 0.6~K/GPa up to 1.6 GPa, for higher pressure
it decreases with a rate of 0.3~K/GPa which is slower than the diminutions of $T_c$ in CeRhIn$_5$.
The very large jump at ambient pressure in the specific heat $\Delta C/C(T_c) = 4.5$, which is the
largest found in heavy fermion superconductors, was first interpreted as a hint for strong coupling
superconductivity in CeCoIn$_5$ \cite{Movshovich01,Sparn01}. The pressure dependence of the the
jump of the specific heat at $T_c$ is shown in figure \ref{figure:CeCoIn5_delta_C}. The height of
the jump obtained by Sparn {\it et al.} was used to determine the background signal for the ac
calorimetry and to normalize the jump obtained by ac calorimetry. The background is of about 40\%
of the measured ac signal in the normal state and it is assumed to be linear in temperature and
independent of pressure.
By increasing pressure the large jump in the specific heat decreases linearly to $\Delta C/C(T_c) =
1$ at 3~GPa. The reduction of the jump with pressure is clearly an indication of the reduction of
the effective mass $m^*$ with increasing pressure. Neglecting strong coupling effects, the jump
of the specific heat normalized to the effective mass $\Delta C/m^* T_c \propto {\rm const.}$
must be fulfilled. However, the weakness of strong coupling is justified by the temperature
variation of the upper critical field of CeCoIn$_5$, which can be expressed in a weak coupling
model with strong Pauli limitation. The large jump at ambient pressure is due to the fact that
superconductivity sets in when the heavy fermion state is not yet formed and the effective mass is
still increasing to lower temperatures. Measurements of the specific heat in field at 5~T parallel
to  $c$ axis show that $C/T$ is increasing to lowest temperatures \cite{Petrovic01b,Bianchi03}. The
increase of $C/T$ is a strong indication that CeCoIn$_5$ at ambient pressure is close to a magnetic
instability and the  system can be driven through a quantum critical point by applying a magnetic
field higher than the upper critical field. By contrast,  at 3~GPa superconductivity sets
in when the heavy masses are formed and CeCoIn$_5$ behaves as a usual heavy fermion system.  The
decrease of the effective mass with pressure has been seen directly by de Haas-van Alphen
experiments under high pressure \cite{Shishido03}.

To estimate the influence of pressure inhomogeneities on the superconducting transition in
CeCoIn$_5$ we calculated the specific heat in the same manner than for CeRhIn$_5$ (see above).
However, in addition the pressure dependence of the effective mass $m^*$ has to taken into account.
The lines in figure \ref{figure:Cac_CeCoIn5_fit} are the results of the calculations. Contrasting
with CeRhIn$_5$,  for CeCoIn$_5$ the pressure distribution increases linearly from $\Delta P =
0.015$ GPa at ambient pressure to $\Delta P = 0.15$ at 2.9~GPa. We can exclude that this increase
of inhomogeneity is due to bare pressure gradients. But it could arise from the material itself. As
pointed out above, uniaxial stress applied in different crystallographic directions may result in
opposite effects on $T_c$, $\partial T_c/\partial \sigma_a > 0$ and $\partial T_c/\partial \sigma_c
< 0$. A stress distribution proportional to the pressure would be the most likely source of this
linear increase of $\Delta P$. The effect of "pressure" inhomogeneities is expected to be more
important in CeCoIn$_5$ than in CeRhIn$_5$, as the anisotropy of the elastic constants of this
compound is the largest of the Ce 115 compounds \cite{Kumar04b}.

\begin{figure}
\begin{center}
\scalebox{0.6}{\epsfbox{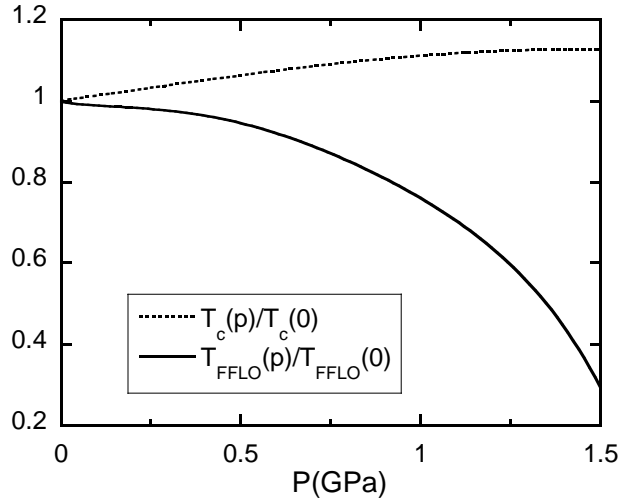}}
\end{center}
\caption{\label{figure:FFLO_p}Relative evolution of $T_c(p)$ and $T_{FFLO}(p)$ as deduced from our
measurements of $Tc(P)$ and of the Sommerfeld specific heat coefficient. Contrary to naive
expectations from the rapid drop of $m^*(P)$, $T_{FFLO}$ is predicted to have only a weak initial
pressure variation.}
\end{figure}

Recently the so-called Fulde, Ferrel, Larkin, Ovchinnikov (FFLO) phase has been found in CeCoIn$_5$
below $T_{FFLO} < T_c$ close to the upper critical field $H_{c2}(0)$ for the magnetic field $H$
applied in the basal plane \cite{Tayama02,Murphy02,Bianchi02,Bianchi03b,Radovan03}. The key point
is that the paramagnetic limit $H_{c2}^p = 1.8 T_c$ in Tesla assuming $g=2$ for the conduction
electrons governs the behaviour of the upper critical field at very low temperatures since the
orbital limit $H_{c2}^{orb}(0)\approx(m^* T_c)^2$ is far higher than $H_{c2}^p(0)$. Nevertheless,
the balance between the orbital and paramagnetic limit is expected to change under pressure, due to
the variation of $T_c$ (which controls $H_{c2}^p$) and of $m^*$, which controls, together with
$T_c$, the $H_{c2}^{orb}$. In fact, both $T_c(P)$ and $m^*(P)$ are known from our experiment, so we
could estimate what should be the relative variation of $T_{FFLO}$ under pressure in a classical
calculation of $H_{c2}$ including the FFLO state (see for example reference \cite{Brison95}). It is
reported on figure \ref{figure:FFLO_p}. From the strong decrease of the effective mass under
pressure, we would have expected a drastic decrease of $T_{FFLO}$ under pressure: this is not the
case, due to the initial increase of $T_c$ which compensates the drop of $m^*$ with $P$. One can
predict that $T_{FFLO}$ will start to decrease significantly only near 1.5~GPa, i.e. when $T_c$
reaches its maximum, despite the fact that at this pressure, $m^*$ has decreased by a factor of 2.
Of course, this prediction is valid only in a classical scheme: it might be different if for
example the interaction itself change with the magnetic field since the FLLO state appears for
magnetic fields just above the field $H_M$ where pseudo-metamagnetism may occur
\cite{Radovan03,Paglione03}.

\section{Conclusion}
We studied the specific heat of CeRhIn$_5$ and CeCoIn$_5$ under high pressure by ac calorimetry and
ac susceptibility up to 3.5~GPa. In CeRhIn$_5$ a first order transition from antiferromagnetic
order below $P_c=1.95$~GPa to a superconducting ground state for $P>2$~GPa has been observed. Below
$P_c$ superconductivity and antiferromagnetism coexist. However, in this regime no superconducting
specific heat anomaly has been observed which points to an inhomogeneous appearance of
superconductivity in this pressure range. Above $P_c$ the very sharp superconducting specific heat
anomaly is due to homogenous bulk superconductivity.

The large jump of the specific heat in CeCoIn$_5$ at the superconducting transition is reduced
linearly with increasing pressure. This is a clear indication for the decrease of the effective
mass with pressure and the system is tuned away from its magnetic instability. At high pressure,
CeCoIn$_5$ behaves like a usual heavy fermion superconductor.

\ack
This work has been supported by the IPMC Grenoble. We thank K.~Miyake for fruitful discussions.


\begin{thebibliography}{99}
\bibitem{footnote}
  As we are only interested in the temperature range very close to $T_c$, the correct form of $C/T$ is not of great importance for the calculation of the broadening of the transition.
\end{thebibliography}
\end{document}